\documentclass[twocolumn,showpacs,preprintnumber,amsmath,amssymb,nofootinbib]{revtex4}

\usepackage{epsfig}
\usepackage{graphicx, subfigure}
\usepackage{epstopdf}

\usepackage[colorlinks]{hyperref}

\usepackage{graphicx}

\usepackage{bm}

\newcommand{\be}{\begin{equation}}
\newcommand{\ee}{\end{equation}}

\newcommand\aap{Astron. Astrophys. }
\newcommand\jcap{J. Cosmol. Astropart. Phys. }
\newcommand\mnras{Mon. Not. R. Astron. Soc. }
\newcommand\apjs{Astrophys. J. Suppl. Ser. }
\newcommand\grg{Gen. Relativ. Gravit. }
\newcommand\etal{{\it et al.}}

\graphicspath{{./figure/}}

\begin{document}
\title{Shear and rotation in Chaplygin cosmology}

\author{A. Del Popolo}
\email{adelpopolo@astro.iag.usp.br}
\affiliation{Dipartimento di Fisica e Astronomia, Universit\`a di Catania, Viale Andrea Doria 6, 95125 Catania, Italy}
\affiliation{Instituto de Astronomia, Geof\'isica e Ci\^encias Atmosf\'ericas, Universidade de S\~ao Paulo, Rua do Mat\~ao 1226, 
05508-900 S\~ao Paulo, S\~ao Paulo, Brazil}


\author{F. Pace}
\email{francesco.pace@port.ac.uk}
\affiliation{Institute of Cosmology and Gravitation, University of Portsmouth,
 Dennis Sciama Building, Portsmouth, PO1 3FX, United Kingdom}

\author{S. P. Maydanyuk}
\email{maidan@kinr.kiev.ua}
\affiliation{Institute for Nuclear Research, National Academy of Sciences of Ukraine,
47 Prospect Nauki, Kiev 03680, Ukraine}

\author{J. A. S. Lima}
\email{limajas@astro.iag.usp.br}
\affiliation{Instituto de Astronomia, Geof\'isica e Ci\^encias Atmosf\'ericas, Universidade de S\~ao Paulo, Rua do Mat\~ao 1226, 
05508-900 S\~ao Paulo, S\~ao Paulo, Brazil}

\author{J. F. Jesus}
\email{jfjesus@itapeva.unesp.br}
\affiliation{Instituto de Astronomia, Geof\'isica e Ci\^encias Atmosf\'ericas, Universidade de S\~ao Paulo, Rua do Mat\~ao 1226, 
05508-900 S\~ao Paulo, S\~ao Paulo, Brazil}
\affiliation{Universidade Estadual Paulista J\'ulio de Mesquita Filho, C\^ampus Experimental de Itapeva,
Rua Geraldo Alckmin 519, 18409-010 Itapeva, S\~ao Paulo, Brazil}

\pacs{}

\date{\footnotesize{Received \today; accepted ?}}

\date{\today}
\begin{abstract}
We study the effect of shear and rotation on results previously obtained dealing with the application of the spherical 
collapse model (SCM) to generalized Chaplygin gas (gCg) dominated universes. The system is composed of baryons and gCg
and the collapse is studied for different values of the parameter $\alpha$ of the gCg. We show that the joint effect of
shear and rotation is that of slowing down the collapse with respect to the simple SCM. This result is of utmost importance for the so-called unified dark matter  models, since the described slow down in the growth of density perturbation can solve one of the main problems of the quoted models, namely the instability 
described in previous papers [e.g., H. B. Sandvik {\it et al.}, Phys. Rev. D {\bf 69}, 123524 (2004)] at the linear perturbation level.

\end{abstract}

\pacs{98.80.-k., 95.36.+x, 95.35.+d}

\maketitle

\section{Introduction} \label{sec:Introduction} 

During the 1990s, numerous results showed that the cold dark matter (CDM) model approach is not
sufficient to describe the observed universe. Nowadays, the scenario that best describes our Universe is a flat
cosmology with dark matter (DM) and an exotic component with a negative pressure, usually named dark energy (DE). This
last component is, in the new picture, the responsible of the accelerated rate of expansion of the Universe. This last
conclusion, coming from the observations of high redshift supernovae, which are dimmer than expectations \cite{SN},
was also confirmed by several others independent observations (e.g. the baryon acoustic {oscillations \cite{Teg},}
the angular spectrum of the CMBR temperature {fluctuations \cite{CM},}
the integrated Sachs-Wolfe {effect \cite{HO}.}
Nevertheless, after a decade of studies, the nature of the DE continues to remain a mystery, and as a consequence 
of this "ignorance", a large number of models have been proposed. The simplest is to identify DE with the 
cosmological constant $\Lambda$, and the energy of vacuum, so obtaining the $\Lambda$CDM model, in which the equation of
state (EoS) of DE is simply given by $w=p/\rho=-1$. In order to alleviate one of the problems of the $\Lambda$CDM model,
namely the cosmological constant concordance problem, several other alternative DE models have been proposed. Extensions
of this model are based on a scalar field weakly interacting with matter (quintessence models) \cite{Gum},
K-essence, phantom models, or unified dark matter models (UDM) (see, e.g., Ref. \cite{AvBe}).
In UDMs, DM and DE are described by the same physical entity. One peculiar case is the so called {\it generalized 
Chaplygin gas} (gCg), introduced by Kamenshchik \cite{Kam}
and then developed in studies by \cite{BBG}.

The EoS describing the gCg is
\begin{equation}
p=-\frac{C}{\rho^{\,\alpha}},
\label{eq:EoS_gCg}
\end{equation}
where $C$ and $\alpha$ are positive constants, $\rho$ is the density, and $p$ is the pressure. 
{When $\alpha = 1$, the} gCg corresponds to the standard Chaplygin gas (sCg)\footnote{The sCg is named
after  Sergey A.\ Chaplygin, the Russian physicist who studied it in a hydrodynamical {context \cite{chaplygin:1901}}}.

Avelino \cite{Ave} showed that the gCg background density evolution is
\begin{equation}
 \rho = \rho_0 \left[\bar C + (1-\bar C)a^{-3(\alpha + 1)} \right]^{\frac{1}{1+\alpha}}\;,
\label{eq:density_gCg}
\end{equation}
where $a$ is the cosmic scale factor, related to the cosmological redshift by $1+z=a_0/a$, and
$\bar{C}=C/\rho_0^{1+\alpha}$, $\rho_0$ is the density at the present epoch.

The EoS parameter, $w$, is given by
\begin{equation}
w = -\bar C\left[\bar C + (1-\bar C)a^{-3(\alpha + 1)} \right]^{-1}\;.
\label{eq:w_gcg}
\end{equation}

It is important to stress that Eq.~(\ref{eq:w_gcg}) shows that the gCg behaves as DM at early time ($a \rightarrow 0$),
 and at later ones $w \rightarrow -1$, approaching a DE behavior.  

Several theoretical \cite{Bord}
and observational consequences of the Chaplygin gas have been studied. Cosmological tests using CMB measurements
\cite{Ben},
measurements of X-ray luminosity of galaxy clusters \cite{Cun},
SNe Ia data \cite{Fab},
lensing statistics \cite{Dev},
have been performed.  

In the context of UDMs with $\alpha \neq 0$, observations of large-scale structure \cite{Mult,avelino:2004}
and comparison of the linear theory with observations have put in evidence some problems of the gCg UDM.
Avelino \cite{avelino:2004} studied the onset of the nonlinear regime in gCg UDMs, showing that the transition from
the DM behavior to the DE one is not smooth, and showed that in gCg UDM non-linear effects generate a non trivial
backreaction in the background dynamics. 
This implies a break down of the linear theory at late times (even on large scales), for all $\alpha \ne 0$ models. 
They also pointed out the need to take into account non-linear effects when comparing with cosmological
observations.

However, notwithstanding the linear perturbation theory has shown that not all $\alpha$ favor structure formation, 
there is a marginal degree of agreement between gCg UDM and large scale structure observations \cite{Bec}.

In order to have a clearer idea of the importance of the gCg as an alternative to the $\Lambda$CDM, it is necessary to
study the non-linear evolution of DM and DE in the Chaplygin gas cosmology. This was partly performed by Ref.
\cite{bilic:2004}. Moreover, a fully non-linear analysis would require SPH simulations (see,
e.g., Refs. \cite{maccio:2004,aghanim:2009,baldi:2010,li:2011}). 
An alternative analytical approach to perform the quoted non-linear analysis and study the non-linear evolution of 
perturbations of DM and DE, is the popular spherical collapse model (SCM) introduced in the seminal paper of Ref.
\cite{GG72}, extended and improved in following papers
\citep{Fillmore1984,Bertschinger1985,Hoffman1985,Ryden1987,Avila-Reese1998,Subramanian2000,Ascasibar2004,Williams2004}. 

The SCM proposed by Ref. \cite{GG72} does not contain non-radial motions and angular momentum. The way to introduce 
angular momentum in the SCM, and its consequences, were studied in several papers \citep{Ryden1987,Gurevich1988a,
Gurevich1988b,White1992,Sikivie1997,Avila-Reese1998,Nusser2001,Hiotelis2002,LeDelliou2003,Ascasibar2004,Williams2004,
Zukin2010,PopGamb}.

Fernandes \cite{Fern} used the SCM to perform the quoted non-linear analysis. Their Friedmann-Lema\^{\i}tre-Robertson-
Walker (FLRW) universe was endowed with two components: gCg  and {\it baryons}\footnote{Radiation was neglected since 
the study considered only the post-recombination epoch}.
An interesting feature of Fernandes' treatment \cite{Fern} is the fact that differently from other works 
(e.g., Refs. \cite{bilic:2004, multamaki:2004, pace:2010}), they considered the collapse of both gCg and \textit{baryons}. 
Moreover, they assumed, for the background and the collapsing region, a time-dependent equation-of-state parameter 
$w$, and derived a more accurate expression for the {\em effective sound speed}, $c_{{\rm eff}}^{2}$, with respect to 
previous studies (e.g., Ref. \cite{abramo:2008}). However, their study did not consider two important factors, namely 
rotation (vorticity), $\omega$, and shear, $\sigma$. Nevertheless, in any proper extension of the SCM the contraction 
effect produced by shear and the expansion one produced by vorticity should be considered, as done by Ref.
\cite{EN2000}. The previous authors studied the effect of shear and vorticity only in DM-dominated universe, and only 
in Ref. \cite{Del} were
shear and vorticity effects considered in the case of DM- and DE-dominated universes. 

In the present paper, we study how the shear, $\sigma$, and the vorticity, $\omega$, change the results obtained by Ref.
\cite{Fern}. 

The paper is organized as follows: Sec.~\ref{sec:SC} summarizes the model used. It reviews the derivation of the
equation of the SCM in presence of shear and vorticity, the {\em effective sound speed} used, and the way equations were
integrated. Sec.~\ref{sec:res} deals with results and Sec.~\ref{sec:conc} with conclusions.

\section{Model} 
\label{sec:SC}

As we already reported, the SCM is a surrogate to N-body simulations to study the evolution of a density perturbation 
in the nonlinear phase. Because of the Birkhoff theorem, a slightly overdense sphere, embedded in the Universe, behaves
exactly as a closed sub-universe. In the model the overdensity is divided into mass shells, each one expanding with the
Hubble flow from an initial comoving radius $x_{\rm i}$ to a maximum one $x_{\rm m}$ (usually named turn-around radius,
$x_{\rm ta}$), and eventually collapse to a singularity (see Refs. \cite{EN2000,Shaw}, to see how the singularity can be
eliminated). 
Collapse to a point will never occur in practice, since dissipative physics and the process of violent relaxation 
will convert the kinetic energy of collapse into random motions, giving rise to a "virialized" structure (virialization 
occurs at $t \approx 2 t_{\rm max}$). Once a non-linear object has formed, it will continue to attract matter in its
neighborhood and its mass will grow by accretion of new material, in the process of "secondary infall".

In the seminal paper of Ref. \cite{GG72},
the authors were interested in the formulation of a theory of infall of matter into clusters of galaxies. The equations 
of dynamics of the structure written by them are relativistic [Eq. (7) of Ref. \cite{GG72}], but they continued the 
treatment thinking in terms of Newtonian mechanics. Their treatment supposed that the structure collapsed radially and 
that non-radial motions were not present. Several following papers showed how to introduce non-radial motions, and
angular momentum, $L$,
\citep{Ryden1987,Gurevich1988a,Gurevich1988b,White1992,Sikivie1997,Avila-Reese1998,Nusser2001,Hiotelis2002,  
LeDelliou2003,Ascasibar2004,Williams2004,Zukin2010} preserving spherical symmetry\footnote{Spherical symmetry is 
preserved if one assumes that the distribution of angular momenta of particles is random, implying a net null mean
angular momentum \cite{White1992}.}. The equations of the SCM with angular momentum can be written as (e.g., Refs.
\citep{Peebles1993,Nusser2001,Zukin2010}):
\begin{equation}
\frac{d^2 R}{d t^2}= -\frac{GM}{R^2} +\frac{L^2}{M^2 R^3}.
\label{eqn:spher}
\end{equation}

SCM equations can be written in terms of the overdensity $\delta$, using General Relativity \cite{Gazt}
or in the Pseudo-Newtonian (PN) approach to cosmology \cite{jsal}.
In the PN approach, the evolution equations of $\delta$ in the non-linear regime has been obtained and used in the 
framework of the spherical and ellipsoidal collapse, and structure formation by Refs.
\cite{Bernardeau1994,Padmanabhan1996,Ohta2003,Ohta2004,Abramo2007}.

In order to obtain the quoted equations, we assume that the fluid satisfies the equation-of-state $P=w\rho$ (we assume
that the velocity of light $c=1$), and we use in the calculation the generalizations of the continuity equation, of 
Euler's equation (both valid for each fluid species $j$), and of Poisson's equation (which is valid for the sum of all
fluids) given {by \cite{jsal,Abramo2007}:}
\begin{equation}
\frac{\partial\rho_{\rm j}}{\partial t}+\vec{\nabla}_{\vec{r}}\cdot\left(\vec{u}_{\rm j}\rho_{\rm j}\right)+
p_{\rm j}\vec{\nabla}_{\vec{r}}\cdot\vec{u}_{\rm j}=0\;,\label{cont-pnc}
\end{equation}
\begin{equation}
\frac{\partial\vec{u}_{\rm j}}{\partial t}+\left(\vec{u}_{\rm j}\cdot\vec{\nabla_{\vec{r}}}\right)\vec{u}_{\rm j}=
-\vec{\nabla}_{\vec{r}}\Phi-\frac{\vec{\nabla}_{\vec{r}}p_{\rm j}}{\rho_{\rm j}+p_{\rm j}}\;,\label{euler-pnc}
\end{equation}
\begin{equation}
\nabla_{\vec{r}}^{2}\Phi=4\pi G\sum_{\rm k}\left(\rho_{\rm k}+3p_{\rm k}\right)\;,\label{poisson-pnc}
\end{equation}
where $\rho_{\rm j}$, $p_{\rm j}$, $\vec{u}_{\rm j}$ and $\Phi$ denote, respectively, the density, pressure, velocity
and the Newtonian gravitational potential of the cosmic fluid. 

It is important at this stage to recall that in order to derive our equations describing the evolution of perturbations, we assume the validity of the pseudo-Newtonian approach. This approach tries to include relativistic effects (like the inclusion of pressure) adding additional terms
to the usual hydrodynamical equations. This is particularly evident in Eqs. (\ref{cont-pnc})-(\ref{poisson-pnc}).
Relativistic contributions play a role in all the equations. Due to the equivalence principle, the pressure now acts as
a source of the gravitational potential in Poisson's equation [Eq. (\ref{poisson-pnc})] and modifies the denominator of
the last term on the rhs of Eq. (\ref{euler-pnc}) (Euler equation). This set of equations is valid only for subhorizon
scales and they do not take into account possible effects at scales larger than the horizon. Despite their limitations,
they proved to be very useful to describe structure formation in quintessence models and their predictions were in
agreement with results of N-body simulations (see for example the appendix in Ref. \cite{pace:2010} and references therein).
The equations as presented here, are not simply a generalization of Newtonian hydrodynamical equations, but they have a
well defined theoretical justification. As detailed in Ref. \cite{pace:2010}, to which we refer for more details, the
pseudo-Newtonian equations can be derived directly from General Relativity, assuming the stress-energy tensor of a
perfect fluid characterized by density and pressure. Given the stress-energy tensor $T^{\mu\nu}$, one computes its 
four-divergence ($\nabla_{\mu}T^{\mu\nu}=0$) and its contraction with the projection tensor
$h_{\alpha\mu}=g_{\alpha\mu}+u_{\alpha}u_{\mu}$. The first, contracted with the four-velocity, gives the general
relativistic continuity equation, the second the Euler equation. Specifying the usual Newtonian metric and under the
assumption of weak field and small velocities ($v\ll c$) we obtain the pseudo-Newtonian equations. Another confirmation
of the validity of our approach is that our perturbed equations (11) and (12) coincide with the set of Eq. (30) in Ref.
\cite{Ma95} assuming further that time derivatives of the gravitational potential are negligible with respect to spatial
derivatives and that perturbations in the pressure term are not adiabatic. It might appear that there are differences
between the two sets of equations, but this is due to the fact that we work 
in the configuration space while Ref. \cite{Ma95} 
worked in the Fourier space. Slightly different is the reasoning behind the derivation of Poisson's equation.
To derive it we can proceed in two different ways. The first one is to combine
Eqs. (23a) and (23d) in Ref. \cite{Ma95},
or work it out directly with the assumed metric.

We would like to recall that in this work, we want to generalize the work of Fernandes \cite{Fern} taking into 
account the contribution of the shear and rotation terms. We therefore closely
follow the derivation of their equations.

Introducing cosmological perturbations in the previous equations, using comoving coordinates, $\vec{x}=\vec{r}/a$,
using $\delta_{\rm j}=\delta\rho_{\rm j}/\rho_{\rm j}$, and assuming that $w_{\rm j}$ and $c_{\rm eff,j}^{2}$ are
functions of time only, the equations for the perturbed quantities are:
\begin{eqnarray}
\dot{\delta}_{\rm j}+3H\left(c_{\rm eff,j}^{2}-w_{\rm j}\right)\delta_{\rm j} &=& \nonumber\\
-\left[1+w_{\rm j}+\left(1+c_{\rm eff,j}^{2}\right)\delta_{\rm j}\right]
\frac{\vec{\nabla}\cdot\vec{v}_{\rm j}}{a}-\frac{\vec{v}_{\rm j}\cdot\vec{\nabla}\delta_{\rm j}}{a},
\label{cont-pert2}
\end{eqnarray}
\begin{equation}
\dot{\vec{v}}_{\rm j}+H\vec{v}_{\rm j}+\frac{\vec{v}_{\rm j}\cdot\vec{\nabla}}{a}\vec{v}_{\rm j}
=-\frac{\vec{\nabla}\phi}{a}-\frac{c_{\rm eff,j}^{2}\vec{\nabla}\delta}
{a\left[1+w_{\rm j}+(1+c_{\rm eff,j}^{2}) \delta_{\rm j}\right]}\;,
\label{euler-pert2}
\end{equation}
\begin{equation}
\frac{\nabla^{2}\phi}{a^{2}}=4\pi G\sum_{\rm k}\rho_{0_{\rm k}}\delta_{\rm k}\left(1+3c_{\rm eff,k}^{2}\right)\;,
\label{poisson-pert2}
\end{equation}
where $c^2_{\rm eff,j}\equiv\delta p_{\rm j}/\delta\rho_{\rm j}$ is the  effective sound speed of each fluid.
\footnote{Note that in the previous equations, $\vec{\nabla}$ refers to gradient with respect to comoving coordinates
$\vec{x}$.}

The previous equations can be simplified as in Ref. \cite{Abramo2007},
\begin{eqnarray}
\dot\delta_{\rm j} & = & -3H(c^2_{\rm eff,j}-w_{\rm j})\delta_{\rm j}
-[1+w_{\rm j}+(1+c^2_{\rm eff,j})\delta_{\rm j}]\frac{\theta_{\rm j}}{a}\label{eq:dot_delta}\;,\\
\dot\theta_{\rm j} & = & -H\theta_{\rm j}-\frac{\theta_{\rm j}^2}{3a}
-4\pi Ga\sum\limits_{\rm k}{\rho_{0\rm {k}}\delta_{\rm k}(1+3c^2_{\rm eff,k})}\;,
\label{eq:dot_theta}
\end{eqnarray}
where $\theta_j \equiv \nabla \cdot \vec{v}_j$ and ${\vec v}_j$ is the peculiar velocity field.

If we do not discard the shear and vorticity, the equations read:
\begin{eqnarray}
\dot\delta_{\rm j} & = & -3H(c^2_{\rm eff,j}-w_{\rm j})\delta_{\rm j} \nonumber\\
& & -[1+w_{\rm j}+(1+c^2_{\rm eff,j})\delta_{\rm j}]\frac{\theta_{\rm j}}{a}\;,
\label{eq:dot_delta1}\\
\dot\theta_{\rm j} & = & -H\theta_{\rm j}-\frac{\theta_{\rm j}^2}{3a} \nonumber\\
& & -4\pi Ga\sum\limits_{\rm k}{\rho_{0\rm k}\delta_{\rm k}(1+3c^2_{\rm eff,k})} \nonumber\\ 
& & -\frac{\sigma_{\rm j}^2-\omega_{\rm j}^2}{a}\;.
\label{eq:dot_theta1}
\end{eqnarray}
In Eq.~(\ref{eq:dot_delta}) the number of equations is equal to the number of cosmological fluid components in the
system. For a `top-hat' profile, resulting in $\vec\nabla\delta_{\rm j}=0$, the peculiar velocity is the same for all
fluids ($\theta_{\rm j}\equiv\theta$, $\sigma_{\rm j}\equiv\sigma$, $\omega_{\rm j}\equiv\omega$), resulting in only one
Eq.~(\ref{eq:dot_theta}) [or Eq.~(\ref{eq:dot_theta1})]. The reason for this is that to preserve the top-hat profile,
all fluids flow in the same way \cite{abramo:2009}.
We remind the reader that shear and vorticity are present already in Eq. (\ref{euler-pert2}), via the term
$(\vec{v}\cdot\vec{\nabla})\vec{v}$. To obtain $\sigma$ and $\omega$, one simply needs to take the divergence of
Eq. (\ref{euler-pert2}).

In terms of the scale factor $a$, and recalling that $\Omega_{\rm j}= \frac{8\pi G}{3H^2}\rho_{\rm j}$, the previous
equations can be expressed in the form: 
\begin{eqnarray}
\delta_{\rm j}^{\prime} & = & -\frac{3}{a}(c^2_{\rm eff,j}-w_{\rm j})\delta_{\rm j}\nonumber\\
& & -[1+w_{\rm j}+(1+c^2_{\rm eff,j})\delta_{\rm j}]\frac{\theta}{a^2H}\;,
\label{eq:dot_delta_a}\\
\theta^{\prime} & = & -\frac{\theta}{a}-\frac{\theta^2}{3a^2H}\nonumber\\
& & -\frac{3H}{2}\sum\limits_{\rm j}{\Omega_{\rm j}\delta_{\rm j}(1+3c^2_{\rm eff,j})}
- \frac{\sigma^2-\omega^2}{a^2 H}\;,
\label{eq:dot_theta_a_omega}
\end{eqnarray}
where the prime denotes the derivative with respect to $a$.

The evaluation of the term $\sigma^2-\omega^2$ was discussed in Ref. \cite{Del}
by defining the ratio $\alpha$ between the rotational and gravitational term in Eq. (\ref{eqn:spher}):
\begin{equation}
\beta=\frac{L^2}{M^3 RG}\;.
\label{eqn:beta}
\end{equation}
In the case of spiral galaxies like the Milky Way $\beta \simeq 0.4$. Its value is larger for smaller size
perturbations (dwarf galaxies size perturbations) and smaller for larger size perturbations (for galaxy clusters the
ratio is of the order of $10^{-6}$).

In order to obtain a value for $\delta_c$ similar to the one obtained by Ref. \cite{ST01},
we set 
$\beta = 0.04$ for galactic masses (see also Ref. \cite{DelPopolo2012}).

In order to integrate Eq.~(\ref{eq:dot_theta_a_omega}) we need to make explicit the $\sigma^{2} - \omega^{2}$ term. 
This was done in Ref. \cite{Del}.

Based on the above outlined argument for rotation, one may calculate the same ratio between the gravitational and the 
extra term appearing in Eq.~(\ref{eq:dot_theta_a_omega}) thereby obtaining
\begin{equation}
\frac{\sigma^2-\omega^2}{a^2 H^2}=-\frac{3}{2}\beta\sum\limits_{\rm j}{\Omega_{\rm j}\delta_{\rm j}(1+3c^2_{\rm eff,j})}\;.
\label{eqn:so}
\end{equation}

As previously told, our Eq.~(\ref{eqn:so}) is based on the assumption that the ratio of acceleration due to the
shear/rotation term to that of the gravitational field, is constant during the collapse [Eq. (\ref{eqn:beta})]. An
objection to this argument could be that angular momentum, $L$, generated by tidal torques could reduce in the
collapsing phase, producing a reduction of the value of $\beta$ and consequently undermining the result of the
calculation. This objection can be disproved as follows. As previously reported, according to the tidal torque theory,
the large scale structure exerts a torque on the forming structure, with the result of imparting angular momentum on the
protohalo 
\cite{Hoyle,Sciama,Peebles,Doroshkevich,White,Wesson,Ryden88,Eisenstein,Catelan96a,Catelan96b}. After the
protostructure decouples from Hubble flow, turns around and starts to collapse, tidal torquing is made almost 
inefficient because the length of the lever arms are reduced (see Fig. 7 in Ref. \cite{Ryden88}) \cite{Catelan96a, 
Catelan96b, Schafer}. Consequently, angular momentum acquisition is maximum at turn-around, and later it remains
constant, since it is not lost in the collapse phase, as discussed by all the previous cited papers, and as known from
the comparison of the galaxies rotation with the tidal torque theory (e.g., Refs. \cite{Catelan96a, Catelan96b, Eisenstein}).
The term $\beta$ is the ratio of the angular momentum acquired through tidal torques (which as told remains constant 
after turn-around), the mass $M$, and radius $R$ of the protostructure. While $M$ remains constant, $R$, is decreasing 
in the collapse, and in the case of a collapsing sphere, its value at virialization is approximately $R_{\rm final}
\simeq 1/2 R_{\rm initial}$. This produces an increase in the term $\beta$. For precision's sake, we should add that in
the protostructure formation two sources of angular momentum are present: a) the angular momentum originated by tidal
torques (about which we spoke till now) connected to bulk streaming motions, and b) the angular momentum originated by
random tangential motions, often refereed to as ``random angular momentum''
\cite{Ryden1987, White1992, Avila-Reese1998, Nusser2001, Ascasibar2004, Williams2004, Ryden88, DelPopolo09,PopKroup}.
Random angular momentum contribute to increase the total value of the angular momentum of the protostructure.

In the following, we will consider $\beta=0.04$ -- corresponding to spiral galaxies similar to the Milky 
Way -- and $\beta=0.02$ and $\beta=0.01$, corresponding to systems in which rotation is less important.

At this point we want to stress out that combining Eqs. (\ref{eq:dot_delta_a}) and (\ref{eq:dot_theta_a_omega}) will lead to
a quite complicated second order ordinary differential equation that generalizes the usual evolution of the
perturbations. To recover it, one has simply to identify $c_{\rm eff}$ with the unperturbed equation of state $w$. It is
also important to notice that our derivation is very general and it should not be considered as an expansion in the
overdensity parameter. The equations obtained have very broad validity and they hold also in the case that $\delta\gg
1$.

The term $c^2_{\rm eff}$ used is the same proposed by Ref. \cite{Fern}, namely:
\begin{eqnarray}
 c_{\rm eff}^2 =
& = &-\frac{C}{\rho^{1+\alpha}}\frac{(1+\delta)^{-\alpha}-1}{\delta} =
w\frac{(1+\delta)^{-\alpha}-1}{\delta}\;.  \label{eq:my_c2eff} \nonumber \\
\label{eq:cef}
\end{eqnarray}
It was obtained by the quoted authors writing $c_{\rm eff}^2$ by using the EoS of the gCg, Eq.~(\ref{eq:EoS_gCg}), and
the relation between the densities in the background and in the collapsed region as follows:
\begin{equation}
 c_{\rm eff}^2=\frac{\delta p}{\delta\rho}=\frac{p_{\rm c}-p}{\rho_{\rm c}-\rho};
\end{equation}
recalling that the perturbed quantities $\rho_{\rm c}$ and $p_{\rm c}$  are related to their background counterparts by:
\begin{eqnarray}
 \rho_{\rm c} & = & \rho+\delta\rho \\
 p_{\rm c} & = & p+\delta p\;,
\end{eqnarray}
and using $\rho_{\rm c} = \rho(1+\delta)$ and Eq.~(\ref{eq:EoS_gCg}).

Eq.~(\ref{eq:cef}) shows that the effective sound speed depends on the collapsed region (through $\delta$) and 
the background (through $w$). The $w$ relative to the collapsed region, namely $w_c$, is given by Eq.~(20) of Ref.
\cite{Fern}:
\begin{equation}
 w_c=-\frac{C}{(\rho(1+\delta))^{1+\alpha}}=\frac{w}{(1+\delta)^{1+\alpha}}. \label{eq:wc_exact}
\end{equation}
In order to study the effect of $\alpha$ on the growth of perturbations, we solved a system of two 
Eqs.~(\ref{eq:dot_delta_a}) (one for gCg and one for baryons), and one Eq.~(\ref{eq:dot_theta_a_omega}). We used 
three values of $\alpha$, namely $\alpha=0$ (model equivalent to the $\Lambda$CDM), and $\alpha=0.5, 1$, and 
$\bar{C}=0.75$. As previously reported, we considered three values for $\beta$, namely $\beta=0.01$, 0.02, and 0.04.
The initial conditions (ICs) for the system, the values of the density parameters, and Hubble constant are the same as in Ref.
\cite{Fern}, and in agreement with recent values for the $\Lambda$CDM \cite{CM}. As {in} Ref. \cite{Fern}, 
$p_{\rm b}=w_{\rm b}=c_{\rm s,b}^2=c_{\rm eff,b}^2=0$.

\section{Results}
\label{sec:res}

In our calculations, we used the same ICs for all the models. In Figs. 1(a)-1(c) [1(d)-1(f)], the solid lines represent the
evolution of $\delta_b$ ($\delta_{gCg}$) in the case the term $\sigma^2-\omega^2$ is not present (similarly to Ref.
\cite{Fern}), and from top to bottom in each plot the values of $\alpha$ changes from $\alpha=1$, 0.5, 0.

\begin{figure*}[!ht]
 \centering
 \includegraphics[angle=0,width=1.0\hsize]{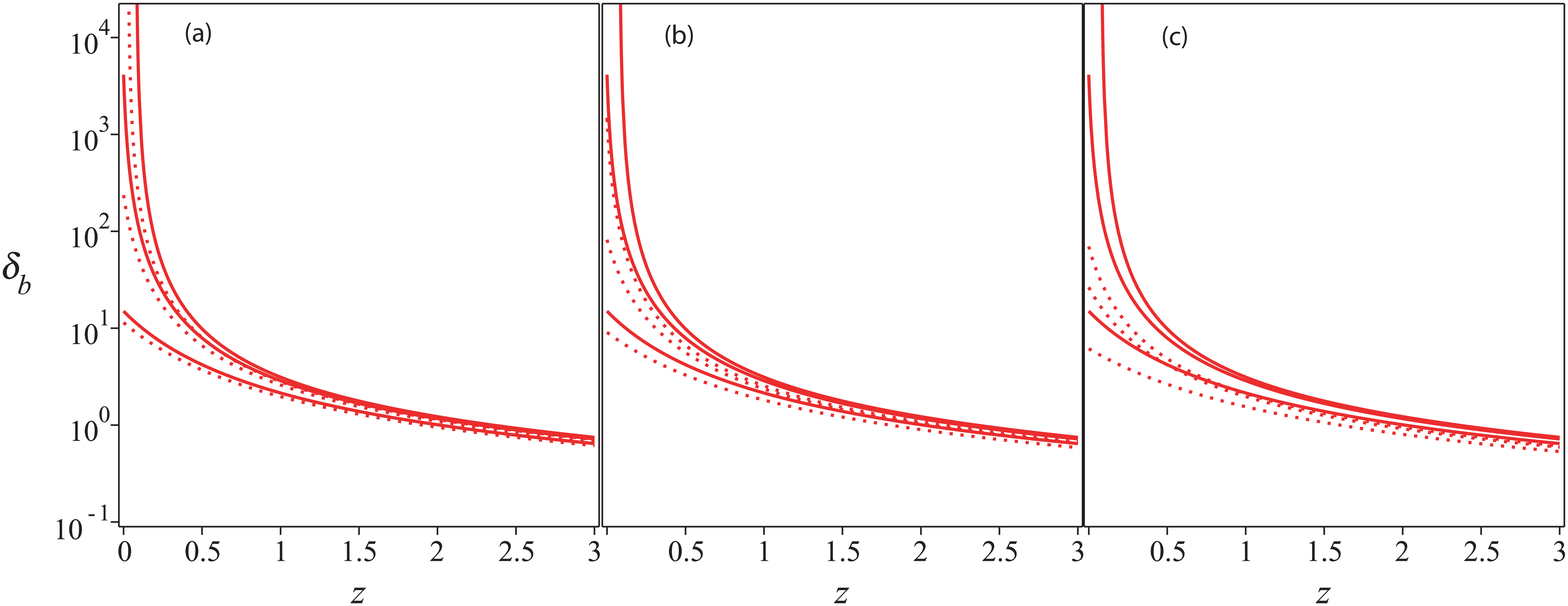}
 \includegraphics[angle=0,width=1.0\hsize]{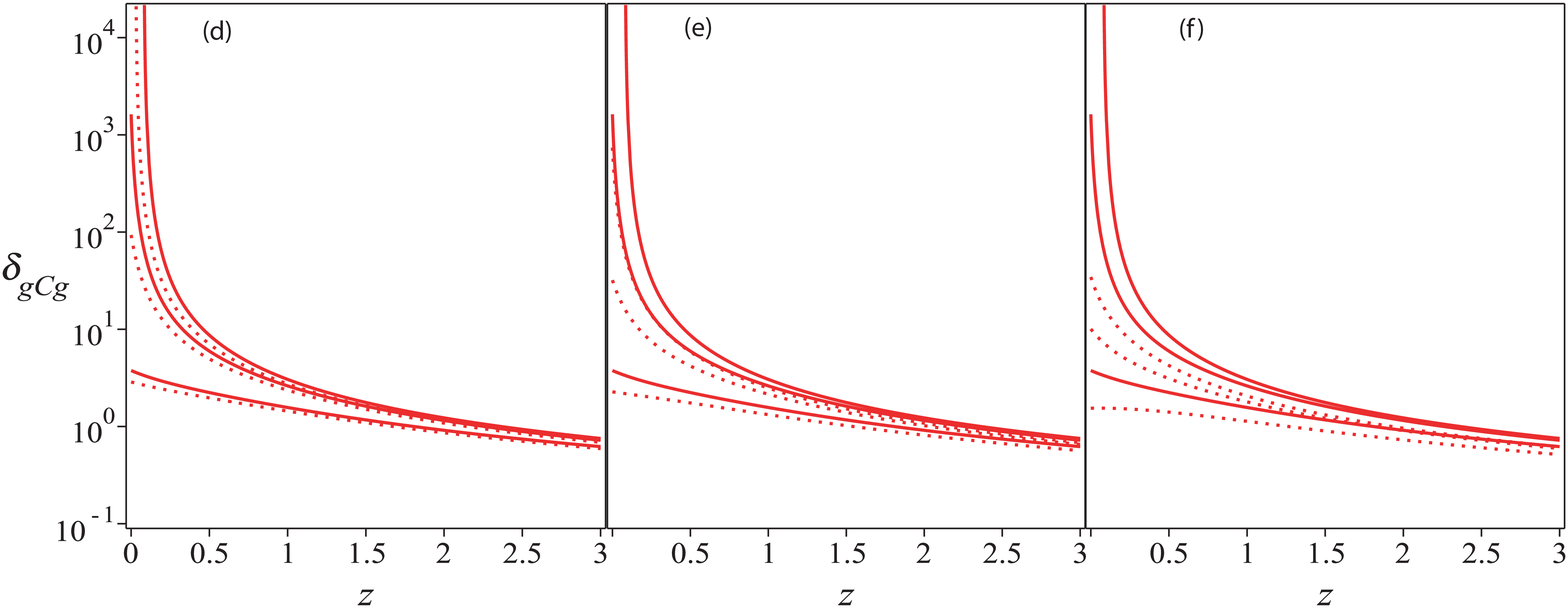}
 \caption{Growth of perturbations for the SCM in gCg-dominated
universes.\textit{Top}: $\delta_b$ vs $z$ (panel a-c). \textit{Bottom}: $\delta_{\rm gCg}$ vs $z$ (panel d-f). Panel
(a-c): from left to right $\alpha$ has the values 1, 0.5, 0. The solid line represents $\delta_{\rm b}$ without the
effect of the additive term, while the dotted line with its presence. The value of $\beta$ is 0.01 in panel a, 0.02 in
panel b, and 0.04 in panel c. Panel (d-f): similar as panel (a-c) but for $\delta_{\rm gCg}$.}
\label{fig:deltas_vs_z}
\end{figure*}

As noticed by Ref. \cite{Fern}, larger values of $\alpha$ produce a faster collapse via larger values of the effective 
sound speed at lower $z$. The different behaviors of $\delta_{\rm b}$ ($\delta_{\rm gCg}$) for different $\alpha$
{reflect} the different evolution of $c^2_{\rm eff}$ and $w$ on the equations of evolution of $\delta$. Moreover, at
smaller $z$ -- when DE dominates -- larger values of $\alpha$ produce a later transition from DM to DE dominated stages of
the gCg universes.

The dotted line represents $\delta_{\rm b}$ ($\delta_{\rm gCg}$) for the same values of $\alpha$ but when
$\sigma^2-\omega^2$ is different from zero. In Fig.~1(a) the value of $\beta$ is 0.01, while in Figs. 1(b) and 1(c) it is 0.02 and
0.04, respectively, and similarly for Figs. 1(d)-1(f). The effect of $\sigma^2-\omega^2$ is that of slowing down the
collapse \cite{DelPopolo02}, so that the collapse acceleration produced by larger values of $\alpha$ is mitigated by the additive term.
Somehow, the presence of the additive term can be mimicked by a reduction of $\alpha$. 

\begin{figure*}[!ht]
 \centering
 \includegraphics[angle=0,width=1.0\hsize]{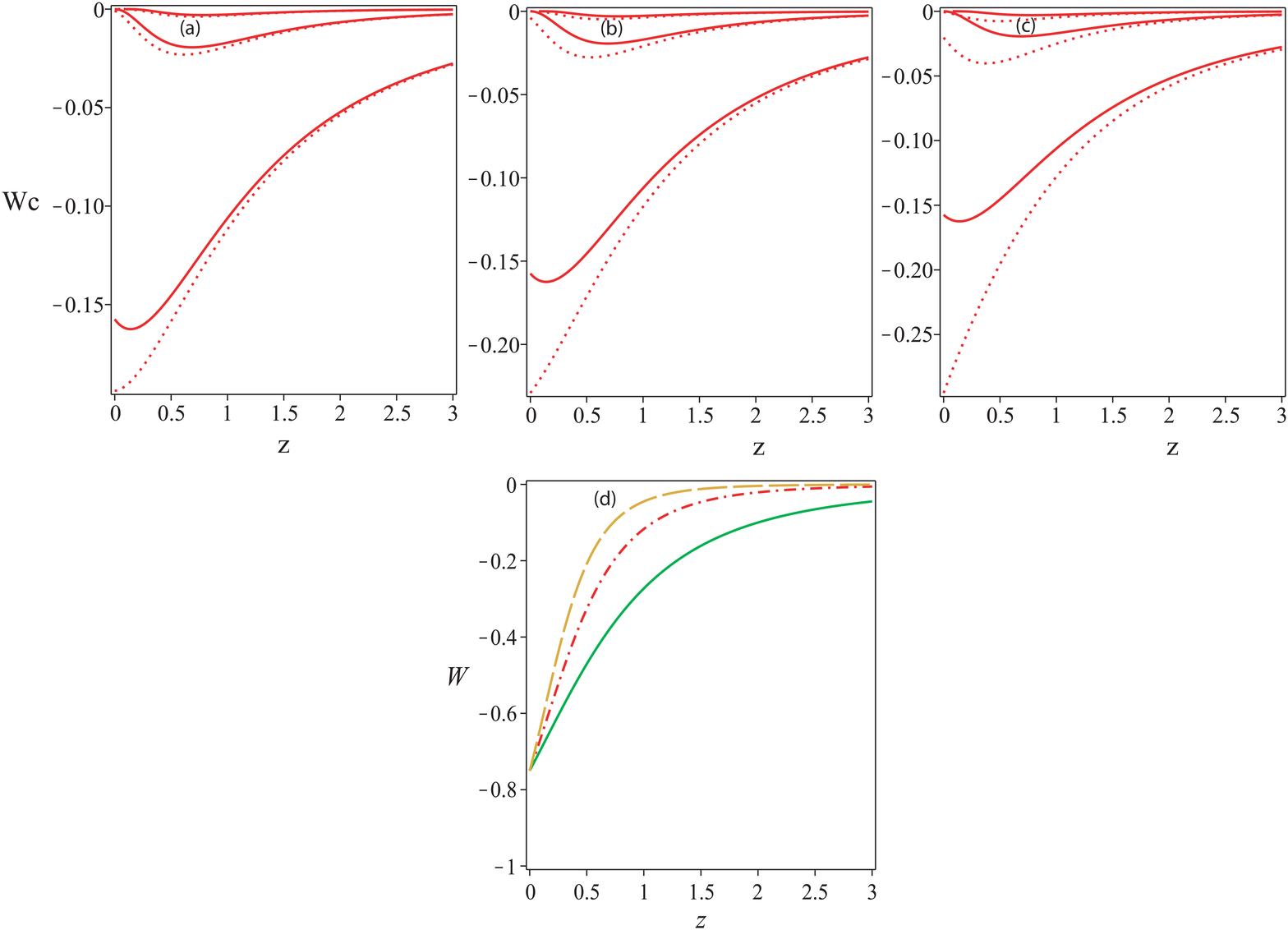}
 \caption{Evolution of $w_c$ and $w$ with $z$ for gCg universes. Panel (a-c): from left to right $\alpha$ goes 
from 1, 0.5, 0. The solid line represents $\delta_{\rm b}$ without the effect of the additive term, while the dotted
line with its presence. The value of $\beta$ is 0.01 in panel a, 0.02 in panel b, and 0.04 in panel c.
Panel (d): $w$ versus $z$ for $\alpha=1$ (dashed line), 0.5 (dot-dashed line), 0 (solid line).}
\label{fig:ws_vz_z}
\end{figure*}

Before going on, we want to stress that a comparison of our calculations with other studies, e.g., Ref. \cite{sandvik:2004},
in order to understand if the instability shown in our Fig 1 (when shear and vorticity are not taken into account) or 
their Fig. 1, is due to a term like $(kc_s/aH)^2\delta_k$, in their Eq. (7), is not trivial. 
A direct comparison of our calculations with theirs [e.g., Eq. (7)] shows that already at linear level, we obtain 
different equations. This is due, as said, to the different approach followed (we followed Fernandes' approach).

In Figs. 2(a)-2(c), we calculate $w_c$ using Eq. (\ref{eq:wc_exact}), and the {previously} calculated values of
$\delta_{\rm gCg}$, while in Fig. 2(d), we calculate $w$. Solid lines in Figs. 2(a)-2(d) represent $w_{\rm c}$ when
$\sigma^2-\omega^2$ is not taken into account, for $\alpha=1$ (top curve), 0.5 (median curve), 0 (bottom curve).
$\alpha$ has a strong effect on the results. Larger $\alpha$ produce a faster collapse and a $w_c$ closer to zero, and
 moreover results in a later transition from DM to DE dominated stages of the gCg universes. The quoted result is obtained 
for a fixed value of $\overline{C}$ ($\overline{C}=0.75$ in our case). If we increase the content of DE of the system, 
corresponding to increasing the value of $\overline{C}$, the collapse will happen at later times or it will be 
prevented with the consequence that $w_{\rm c}$ will no longer be close to zero. The dotted lines represent the same 
quantity when the additive term is taken into account. In Fig. 2(a) the value of $\beta$ is 0.01, while in Figs. 2(b) and 2(c)
it is 0.02 and 0.04, respectively. Since the effect of the additive term is to slow down the collapse, the effect on
$w_{\rm c}$ is that of showing a more pronounced departure from zero. Fig. 2(d) represents $w$ given by
Eq.~(\ref{eq:w_gcg}), depending on $\overline{C}$, $a$, and $\alpha$, and then independent from our additive 
parameter, and consequently identical to Fig. 2(b) of Ref. \cite{Fern}. The solid line represents the case $\alpha=0$, the 
dash-dotted one the case $\alpha=0.5$, and the dashed line the case $\alpha=1$. 

\begin{figure*}[!ht]
 \centering
 \includegraphics[angle=0,width=1.0\hsize]{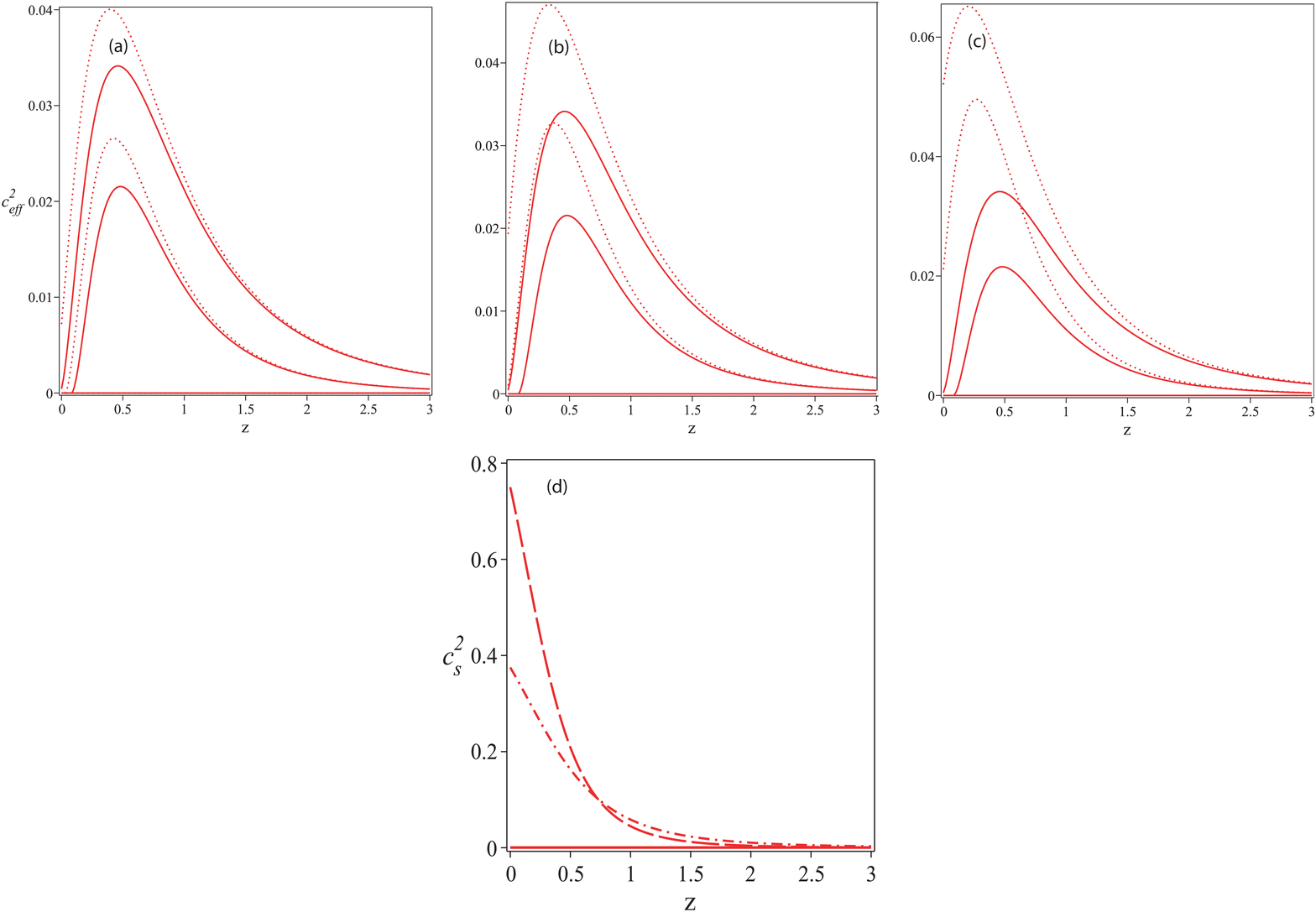}
 \caption{Evolution of $c_{{\rm eff}}^{2}$ and $c_s^2$ with $z$ for gCg universes. Panel (a-c): from left to right 
$\alpha$ goes from 1, 0.5, 0. The solid line represents $c_{{\rm eff}}^{2}$ without the effect of the additive term,
while the dotted line with its presence. The value of $\beta$ is 0.01 in panel a, 0.02 in panel b, and 0.04 in panel c. 
Panel (d): $c_s^2$ versus $z$ for $\alpha=1$ (dashed line), 0.5 (dot-dashed line), 0 (solid line).
}
\label{fig:ceff2_vz_z}
\end{figure*}

\begin{figure*}[!ht]
 \centering
 \includegraphics[angle=0,width=1.05\hsize]{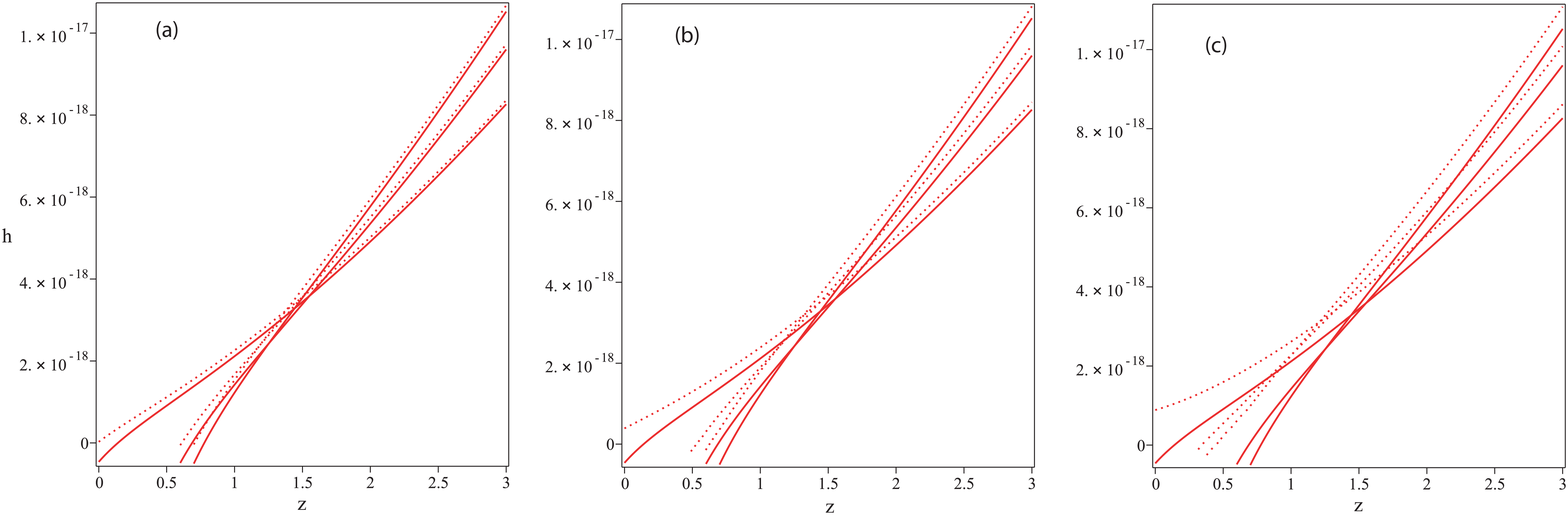}
 \caption{Evolution of the expansion rate, $h$, with $z$. Solid lines, represent the case  $\sigma^2-\omega^2=0$. 
From left to right, in each panel: $\alpha=0$, 0.5, and 1. Dotted lines: $\sigma^2-\omega^2 \neq 0$, with $\beta=0.01$
(Fig. 4a), $\beta=0.02$ (Fig. 4b), and $\beta=0.04$ (Fig. 4c).
}
\end{figure*}

In Figs. 3(a)-(c) we plot $c^2_{\rm eff}$ and in Fig. 3(d) $c_{\rm s}^2=-\alpha w$ for the same values of $\alpha$ and
$\beta$, and with the lines having the same meaning as in previous figures. The plot shows the different behavior of
$c^2_{\rm eff}$ and $c_{\rm s}^2$, implying a different behavior of the gCg component locally ($c^2_{\rm eff}$) and in
the background ($c_{\rm s}^2$). Again notice that our Fig. 3(d) for $c_s^2$, is the same of Fig.~3(b) of Ref. \cite{Fern}, since
the sound speed is not dependent upon the additive parameter.

Finally, Fig.~4 plots the evolution of $h=H+\frac{\theta}{3a}$ with $z$. Solid lines again represent $h(z)$ without 
the additive term. In each plot, $\alpha$ has the values $\alpha=0$, 0.5, 1, from left to right. Larger values of
$\alpha$ give rise to a faster decrease in $h$. Since the turn-around redshift, $z_{\rm ta}$, can be defined as the $z$
at which $h=0$, it is clear that higher $\alpha$ imply a larger $z_{\rm ta}$ and an earlier collapse. Taking into
account the additive term, $z_{\rm ta}$ becomes smaller with respect to the case in which it is not present. The effect
of the term $\sigma^2-\omega^2$, is represented by the dotted lines: $\beta=0.01$, $0.02$, and $0.04$ in Figs. 4(a)-4(c), respectively.

Previous works (e.g., Refs. \cite{sandvik:2004,gorini:2008}) have shown a problem in UDM models, namely oscillations or exponential
blowup of the dark matter power spectrum not seen in observations -- a problem which is evident on galactic scales and only
at recent times, and that cannot be solved by taking baryons into account, as proposed by Ref. \cite{Col}.
Both Refs. \cite{sandvik:2004,beca}, have shown that gravitational effects of DM, at late time, can add fluctuations 
to baryons but that they are unable to erase the ones already present.

Our result concerning the effect of $\alpha$ on the growth of perturbations are partially in disagreement with the  
linear theory of perturbation in gCg universes (e.g., Refs. \cite{sandvik:2004,gorini:2008}), and in agreement  
with the findings of Ref. \cite{Fern}. However, in our study -- due to the additive term, which has its maximum effect on galactic 
scales -- the growth of perturbation is slowed down. This somehow implies that the additive term presence 
works in the direction of reducing possible present oscillations as found by Refs. \cite{sandvik:2004,gorini:2008}.

The previous results are perfectly framed in top-hat profiles for density and pressure. 
Since the profile is flat, it does not contain pressure gradients and the growth of perturbations can be suppressed only 
by an accelerated expansion. Assuming a nonflat initial perturbation, it would be possible to improve the understanding
of how $\alpha$ affects structure formation.

The study of the effect of pressure gradients, by using alternative profiles -- such as a Gaussian profile or a 
Navarro-Frenk-White profile -- will be the object of a future study.

Similarly to Ref. \cite{Fern}, we also studied how changing the ICs changes the turn-around epoch. Small changes of the ICs
can produce the same turn-around redshift in all models.

\section{Conclusions}
\label{sec:conc}

In the present paper, we used the SCM to study how perturbations evolve in gCg universes, taking into account the 
effect of shear and rotation. We used the same $c_{\rm eff}^2$ as Ref. \cite{Fern}, and in agreement with their work we found
that larger values of the parameter $\alpha$ speed up the collapse, but the additive term $\sigma^2-\omega^2$ produces a
slowing down of this acceleration, visible in the figures showing the evolution of $\delta_{\rm b}$, $\delta_{\rm gCg}$
(Fig. 1), $w_{\rm c}$, $c_{{\rm eff}}^{2}$, and $h$. The comparison of $w_{\rm c}$, and $c_{\rm eff}^2$ (the local,
nonlinear parameters) with $w$, and $c_{\rm s}^2$ (the global, linear parameters) shows clear evidence of the difference in the
linear and nonlinear dynamical behavior of the gCg.   

Notwithstanding that the SCM is usually a faithful technique to study gravitational collapse and structure formation --   
with results comparable to those of simulations \cite{ascasi} --
it would be worthwhile to check the results of this paper against SPH simulations, that would allow to take account of
spatial pressure gradients. Moreover, more realistic profiles would improve our understanding of the local dynamics 
of {gCg} universes, and how the background dynamics is influenced by local non-linear inhomogeneities. 

\vspace{0.05cm}
\vspace{0.1cm}{Acknowledgments:} 
A.D.P. is partially supported by a visiting research fellowship from FAPESP (Grant No. 2011/20688-1), and also wishes to thank 
the Astronomy Department of S\~ao Paulo University for the facilities and hospitality. F.P. is supported by STFC Grant No.
ST/H002774/1, J.F.J. is supported by FAPESP Grant No. 2010/05416-2, and J.A.S.L. is also partially supported by CNPq and FAPESP
under Grants No. 304792/2003-9 and No. 04/13668-0.

\end{document}